\documentclass[11pt]{article}

\usepackage[final]{acl}

\usepackage{times}
\usepackage{latexsym}

\usepackage[T1]{fontenc}
\usepackage[utf8]{inputenc}

\usepackage{microtype}

\usepackage{graphicx}
\usepackage{booktabs}
\usepackage{hyperref}

\definecolor{darkblue}{rgb}{0, 0, 0.5}
\hypersetup{colorlinks=true, citecolor=darkblue, linkcolor=darkblue, urlcolor=darkblue}

\usepackage{amsfonts}
\usepackage{amssymb}
\usepackage{amsmath}
\usepackage{nicefrac}
\usepackage{xcolor}
\usepackage{float}
\usepackage{multirow}
\usepackage{makecell}
\usepackage{bm}
\usepackage{subcaption}
\usepackage{tcolorbox}
\tcbuselibrary{breakable, skins}
\usepackage{listings}
\usepackage{stfloats}
\fnbelowfloat
\usepackage{xurl}

\definecolor{commentgreen}{RGB}{2,112,10}
\definecolor{eminence}{RGB}{108,48,130}
\definecolor{weborange}{RGB}{255,165,0}
\definecolor{frenchplum}{RGB}{129,20,83}

\newcommand{\code}[1]{\texttt{#1}} 

\title{Generation and Verification: Human-Agent Collaboration for Safety-Critical Automotive Software Engineering}

\author{
  \textbf{Elizaveta Zinovyeva\textsuperscript{1}},
  \textbf{Aiham Taleb\textsuperscript{1}},
  \textbf{Nikita Kozodoi\textsuperscript{1}},
  \textbf{Daniel Schleicher\textsuperscript{1}},
\\
  \textbf{Amir Mahdi Namazi\textsuperscript{2}},
  \textbf{Martin Kraus\textsuperscript{2}},
  \textbf{Levent Kent\textsuperscript{1}},
  \textbf{Brian Jensen\textsuperscript{1}\thanks{Work done while at Amazon Web Services.}}
\\
\\
  \textsuperscript{1}Amazon Web Services, Germany,
  \textsuperscript{2}AUMOVIO, Regensburg, Germany
\\
  \small{
    \texttt{\{zinovyee,aitaleb,kozodoi,dschlei,levkent,jensbri\}@amazon.com}
  }
\\
  \small{
    \texttt{\{amir.mahdi.namazi,martin.2.kraus\}@aumovio.com}
  }
}
\begin{document}
\maketitle

\begin{abstract}
Research on coding agents mainly focuses on generation quality. Yet, safety-critical automotive software makes two additional demands. It prioritizes verification, since every change is reviewed, traced to a requirement, and checked against industry standards such as MISRA and AUTOSAR before it can be committed. It is also a narrow domain, written against layered architectures, embedded resource constraints, and organization-specific interface conventions that are scarce in the data general-purpose LLMs are trained on. We present a coding assistant for automotive C/C++ that addresses both demands. First, it exposes three tools to the user over the Model Context Protocol (\code{generate}, \code{edit}, and \code{verify}), allowing developers to run \code{verify} on any code on demand rather than relying on the agent to check its own output. Second, rather than using a general-purpose model, we customize the underlying LLM for the automotive domain through prompt engineering and fine-tuning. We construct a dataset from automotive production code of AUMOVIO, where each function is paired with the requirement texts and Unified Modeling Language (UML) behavior it must satisfy. In a two-week pilot on active production projects, developers report \code{verify} as the most-used tool and confirm using it on their own code and not only on generated output. Comparing model customization strategies, 75\% prefer the faster prompt-engineered configuration. MISRA- and AUTOSAR-inspired pattern checkers indicate that prompt engineering achieves a lower number of violations on C, while fine-tuning shows better conformance with architecture patterns for C++.
\end{abstract}

\section{Introduction}\label{sec:introduction}
Research on AI coding assistants is organized largely around generation quality \citep{jiang2024survey}, asking whether a model produces code that compiles, passes tests, and resembles a reference implementation \citep{paul2024benchmarks,zheng2024beyond}. Automotive software development is adopting these assistants as car makers seek productivity gains \citep{automotive2025mckinseay}. Yet, they fall short of safety-critical requirements such as AUTOSAR \citep{autosar2017guidelines} and MISRA-C \citep{motor2013misra}, which establish strict rules for reliability in automotive systems. Recent studies show that popular LLMs struggle to produce MISRA-compliant code by default \citep{umer2025comparative}, while hallucinations pose tangible safety risks \citep{pavel2025hallucination}. This places two additional demands on coding assistants for automotive software.

The first demand is verification. In safety-critical settings, developers cannot accept LLM output blindly, since every change is reviewed, traced to a requirement, and checked against coding standards before it can be committed. The practical question is therefore not only whether the model can write compliant code, but how the agent should collaborate with the developer so that compliance can be verified, audited, and trusted.

The second demand is domain fit. Automotive code must guarantee deterministic execution under real-time constraints, follow specific architectural patterns (e.g., AUTOSAR layered architecture) rarely encountered during pre-training, operate under tight memory and resource constraints in embedded systems, and respect interface conventions particular to the organization that owns the codebase. A general-purpose model fits this setting poorly, since the conventions it must follow are underrepresented in its training data \citep{cassano2024knowledge,gu2025effectiveness,liu2023codegen4libs}.

This paper introduces an agentic assistant for automotive software, built on the Model Context Protocol (MCP) \citep{mcp2024anthropic}, that addresses both demands. We address the first demand by exposing verification as a separate tool. Rather than providing a single code generation mode, the system exposes three tools: \code{generate}, \code{edit}, and \code{verify}. The first two generate and revise function bodies from a signature, documentation, and context such as imported functions and UML diagrams. This task is central to automotive workflows, where interfaces are specified upfront and implementations must follow established patterns. The third tool lets developers run the same automotive rule checks the agent uses internally, on demand and independently of generation. Most coding agents treat checking as an internal step that the agent itself decides to invoke. We argue that in safety-critical settings the developer must be able to trigger evaluation directly, because a check the agent can skip is not a reliable check.

We address the second demand by customizing the model to the domain. We build a dataset from an AUMOVIO's production automotive code in which each function is paired with the context needed to implement it compliantly, including requirement texts resolved from AUMOVIO's requirements-management system and the UML behavior the function must realize. On this dataset we compare the two customization strategies most used for code models \citep{shin2025prompt,freire2026exploring}: prompt engineering (PE) with general-purpose API models and domain-specific fine-tuning (FT) \citep{zirak2024improving}, both scored against custom rule sets inspired by established automotive standards. PE here involves supplying the assembled context and the rule sets in full at inference time, rather than tuning instruction wording, so the two strategies differ in when the domain information reaches the model and not in what it contains. Because those rule sets are pattern matchers, we treat their output as pattern conformance rather than compliance, and report how that changes what the numbers support.

The main contributions are the following. First, we present an MCP-based architecture in which verification is exposed as a tool the developer can invoke directly, on any code and at any time, and articulate the corresponding division of labor between human and agent in safety-critical review. Second, we implement a context engineering pipeline that assembles requirement texts, UML behavior, and API context into training and inference examples, which lets us compare PE against FT on the same domain information, differing in whether it arrives during training or at inference. We analyze how the two methods perform in terms of architecture and coding pattern violations and generation latency. Comparing against a human baseline of shipped production code, we find that generated C code scores better than code written by the engineers, and we show that this reflects what the checkers reward rather than a genuine quality advantage. Finally, we report results from a two-week pilot with nine developers from AUMOVIO and show how the three tools are taken up in a real-world production workflow.

\section{Related Work}\label{sec:rel_work}
Code generation and editing are part of the Machine Learning on Code or Big Code research field \citep{allamanis2018survey}. Code generation for automotive software dates back to the late 1980s and 1990s \citep{hanselmann1998development,hanselmann1999automatic}, where early systems used rule-based transformations; neural approaches have become feasible only recently with the advent of LLMs. Early neural code generation models such as Codex \citep{chen2021evaluating} and open-source counterparts demonstrated strong accuracy in general programming~\citep{li2022alphacode,wang2021codet5,ahmad2021plbart}, but struggle with domain-specific code due to gaps in specialized knowledge. \citet{gu2025effectiveness} show that general-purpose LLMs underperform in domain-specific contexts because they make limited use of domain-specific libraries, and their DomCoder approach recovers quality by injecting API knowledge through prompting and fine-tuning. Another work fine-tunes code models on domain-specific corpora \citep{liu2023codegen4libs} and finds that developers expect assistants to understand specific frameworks and libraries, and their two-stage CodeGen4Libs outperforms non-domain-specific baselines.

Which of the two customization strategies to prefer is unsettled. \citet{shin2025prompt} compare prompt engineering against 17 fine-tuned models on code summarization, generation, and translation, and find that prompting a stronger general-purpose model does not consistently win, while \citet{freire2026exploring} report that LoRA fine-tuning beats few-shot prompting and retrieval on domain-specific text-to-code. This motivates our comparison of the two strategies on the same assembled context.

For safety-critical domains, researchers integrate verification and domain constraints into generation. \citet{fakih2024llm4plc}'s LLM4PLC combines fine-tuning with an iterative verification loop, improving code generation success rates from 47\% to 72\% and expert-rated quality from 2.25/10 to 7.75/10, while \citet{jiang2025rocode}'s ROCODE integrates static analysis to reach a 99.1\% compilation success rate. Recent work targets automotive compliance specifically: \citet{kirchner2025generating} integrate LLMs into automotive development with test-driven development and iterative refinement; \citet{umer2025comparative} finds that popular LLMs (ChatGPT, Gemini, DeepSeek, Copilot) produce no MISRA C++ compliant code by default; and \citet{pavel2025hallucination} identify high rates of syntax violations and API conflicts in GPT-4 variants that pose safety risks. Together, these indicate that domain-specific code generation requires both LLM adaptation and domain-aware validation. Such validation is typically placed inside the generation loop, where the model critiques and repairs its own output, but gains from self-repair are modest and bottlenecked by the model's ability to judge its own code \citep{olausson2024selfrepair}, and models struggle to self-correct without external feedback \citep{huang2024selfcorrect,stechly2024selfverification}.

A further stream studies coding assistants from a developer-productivity perspective, reporting faster task completion with GitHub Copilot \citep{peng2023copilot} and measuring suggestion acceptance and satisfaction across a 400-developer enterprise rollout \citep{bakal2025zoominfo}. \citet{heydari2025humancentric} argues that such evaluations still under-weigh how well a tool adapts to user expertise and supports collaboration. Productivity and adoption therefore depend not only on generation quality but on how the assistant supports oversight, which is amplified in safety-critical settings.

A recent line of work reframes review of generated code as a trust problem rather than a diffing problem. \citet{heander2026trustcalibrated} find, across a participatory design study with industry practitioners, that trust calibration is the central challenge when reviewing agent-generated multi-file changes, and argue that tools should surface risk and confidence signals at the granularity where reviewers allocate attention. \citet{mitropoulos2026contextual} supply the security argument for keeping the reviewer independent of the generator: automated review judgments are systematically swayed by how a change is framed, and an iterative attack that refines its framing against a local copy of the review pipeline succeeds in every case they test, since the attacker may retry while the defender has one chance to detect. Both motivate our design: verification the developer controls, rather than verification the generating system elects to perform. Our work differs from these lines in treating verification as a developer-invocable tool on par with generation, grounding the comparison in production code, and a pilot on an active automotive project rather than synthetic benchmarks.

\section{Methodology}\label{sec:methodology}
Our coding assistant centers on an MCP-based framework compatible with modern AI coding assistants such as Cline, Claude Code and Codex, extending them with automotive-specific tools. We compare two ways of customizing the underlying model: prompt engineering (PE), which supplies domain knowledge at inference time, and fine-tuning (FT), which absorbs it into the parameters. We evaluate \textbf{Qwen3-32B} \citep{yang2025qwen3technicalreport} in three variants, PE, FT, and FT on curated redundancy-removed data, and use \textbf{Claude Sonnet 4} and \textbf{Amazon Nova Pro} in PE mode as baselines.

\subsection{Dataset Preparation}\label{subsec:dp}

\begin{table*}[b]\centering
\begin{lstlisting}[language=C, caption={Abbreviated training example. Each numbered block is assembled by the pipeline from a different artifact in AUMOVIO's repositories; requirement texts are resolved from DOORS identifiers cited in the Doxygen comment. Appendix~\ref{app:format} gives the full untruncated example.}, label=code_example, basicstyle=\ttfamily\scriptsize, xleftmargin=2.5em, framexleftmargin=2.5em]
INPUT CONTEXT
 (1) Signature:  void ValidateEngineParameters(
         const EngineData_t* engineData, ValidationResult_t* result)
 (2) Doxygen:    @brief Validates engine parameters ...
                 @requirements REQ_ENG_001, REQ_ENG_002   @safety_level ASIL-D
 (3) Requirements (resolved from DOORS):
       REQ_ENG_001: "... shall validate that engine RPM remains within the
                     operational range of 800 to 6000 RPM ..."
 (4) UML (PlantUML, if referenced):
       ValidateEngineParameters -> CheckRPMRange : validate RPM
 (5) API context: signatures + docs of imported functions, header declarations
 (6) Rules (optional): JSON rule sets, see Listing &\ref{RULING}&

OUTPUT
 if ((engineData == NULL) || (result == NULL)) { ... return; }   /* null guard */
 if ((engineData->rpm < 800U) || (engineData->rpm > 6000U))      /* REQ_ENG_001 */
 { result->errorCode = VALIDATION_RPM_OUT_OF_RANGE; return; }
\end{lstlisting}
\end{table*}

\noindent
We build a context engineering pipeline to process an AUMOVIO's production automotive C/C++ code into a structured dataset for fine-tuning and evaluation. The data comes from multiple actively maintained projects, so the examples reflect industry practices and compliance requirements. The pipeline parses C/C++ files to isolate individual functions, then augments each with the context a developer would consult when implementing it: header declarations and the signatures and documentation of imported functions for API context, import statements and namespaces, and the requirement and UML artifacts described next. It keeps only functions whose documentation is complete enough to specify behavior, validates them against our custom rule sets, and analyzes complexity for balanced stratification. The pipeline yields over 6,000 function implementations stratified by complexity, of which 822 are held out for evaluation.

The step that distinguishes this setting from general code generation is requirements traceability. In safety-critical automotive practice, an implementation is not merely expected to work; it must be attributable to a specific requirement, and that link is what makes review and certification possible. AUMOVIO's code carries this link in a machine-readable form: Doxygen comments cite requirement identifiers, while the requirement texts live in a separate requirements-management system. Our pipeline parses DOORS \citep{ibm_doors} export XML to expand each cited identifier into its full requirement text, and likewise extracts referenced PlantUML \citep{plantuml} diagrams to supply behavioral specifications. The pipeline also reconciles implementation signatures against their header declarations, correcting mismatches that would otherwise train the model on inconsistent interfaces. Each function's prompt therefore states the complete obligation that the function is meant to satisfy. 

Listing~\ref{code_example} shows the resulting format of the constructed function prompt, with the requirement text and UML fragment resolved in place. Appendix~\ref{app:preprocessing} details the extraction and filtering steps.

\subsection{Model Customization}

In PE mode, model parameters stay fixed and all domain knowledge enters through the prompt. Each request carries a system prompt that states the automotive coding role and instructs the model to return a single function implementation based on the task context assembled by our pipeline: the function signature, its documentation and linked requirements, signatures and documentation of imported functions, and the UML diagram when one is referenced. The prompt also includes JSON rule sets the code needs to conform with, including conforming and non-conforming examples (see Section~\ref{subsec:rules} for details). The rule sets are supplied in full rather than retrieved per request, since our subset is compact enough to fit the context window of all models we test, which keeps the prompt identical across examples and avoids confounding retrieval quality with model capability. Because the assembled context is the same one used to build training examples, the two strategies work from the same domain information, differing in whether it arrives at inference time or during training.

In FT mode, the same information is instead absorbed into the parameters. We select Qwen3-32B as the base model and train it on the preprocessed function corpus formatted as user-assistant turns via HuggingFace chat templates, with the context as the user message and the human production implementation as the target, so the model learns the conventions its authors applied rather than a synthetic notion of compliant code. We use Low-Rank Adaptation \citep{hu2021loralowrankadaptationlarge} with rank-8 adapters to adapt the model while preserving general coding capabilities, with 4-bit quantization \citep{dettmers2023qlora} and gradient checkpointing to fit available memory, on SageMaker with DeepSpeed \citep{rasley2020deepspeed} for multi-GPU training (Appendix~\ref{app:training}). We train two variants: one on the full corpus and one on a curated version with near-duplicate functions removed. The resulting models are packaged through Amazon Bedrock's custom model import functionality for integration with the multi-model MCP architecture. At inference, the FT models receive the same assembled context and rule sets as the PE models, so FT adds parameter adaptation on top of the PE condition.

\begin{table*}[b]\centering
\begin{lstlisting}[language=C, caption=Simplified C Example: Buffer Access Rule, label=RULING, xleftmargin=2.5em, framexleftmargin=2.5em]
{"rule_id": "BUF_001",
"rule_name": "Buffer Access Protection",
"non_compliant_example": "memcpy(dest, src, len);",
"compliant_example":
"if (dest && src && len > 0 && len <= DEST_SIZE && len <= src_size)
{
memcpy(dest, src, len); return SUCCESS;
}
return ERROR;"}
\end{lstlisting}
\end{table*}

\subsection{Automotive Rule Sets Integration}\label{subsec:rules}

We build custom rule sets inspired by established automotive coding standards, covering patterns common in safety-critical embedded systems and prioritized by AUMOVIO's engineers. 

The \textbf{Coding} rules, 57 for C and 107 for C++, address pointer validation before dereferencing, buffer bounds checking, error return handling, and memory access patterns. The \textbf{Architecture} rules, 37 for C and 30 for C++, address layered architecture principles, interface design conventions, and module separation patterns typical in automotive software. Each rule carries one of four severity levels (info, warning, error, critical), mirroring the tiers used in static analysis tools. We review the rule sets with automotive engineers to confirm they target the patterns AUMOVIO prioritizes. The same rule sets serve two purposes: they supply context during code generation, and they act as scoring rubrics during evaluation (Section~\ref{sec:evaluation}).

We process the rule sets into a JSON format (see Listing~\ref{RULING}) holding rule identifiers, descriptions, and pairs of conforming and non-conforming examples.

\subsection{Deployment}

We package the solution as an MCP-based agentic coding assistant with three core tools, each mapping to a stage of the developer's workflow: \code{generate} produces new function implementations aware of custom rule sets from a specification or signature; \code{edit} makes contextual modifications using existing code for refactoring or extending; \code{verify} checks code against the rule sets.

In the standard MCP arrangement, an agent reads tool descriptions and selects tools autonomously, so a well-described checker is invoked whenever the agent judges a check useful. We retain this setup but also expose \code{verify} to developers directly. The concern is not that agents skip checks; it is that the model which produced a piece of code is poorly placed to judge whether that code warrants scrutiny. Safety-critical review practice resolves this through independence, where a different party controls the check. Exposing \code{verify} as a peer tool the developer invokes on demand preserves that independence.

\begin{figure*}[t] \centering
\includegraphics[trim={0 0 0 0}, width = 0.75\textwidth,clip]{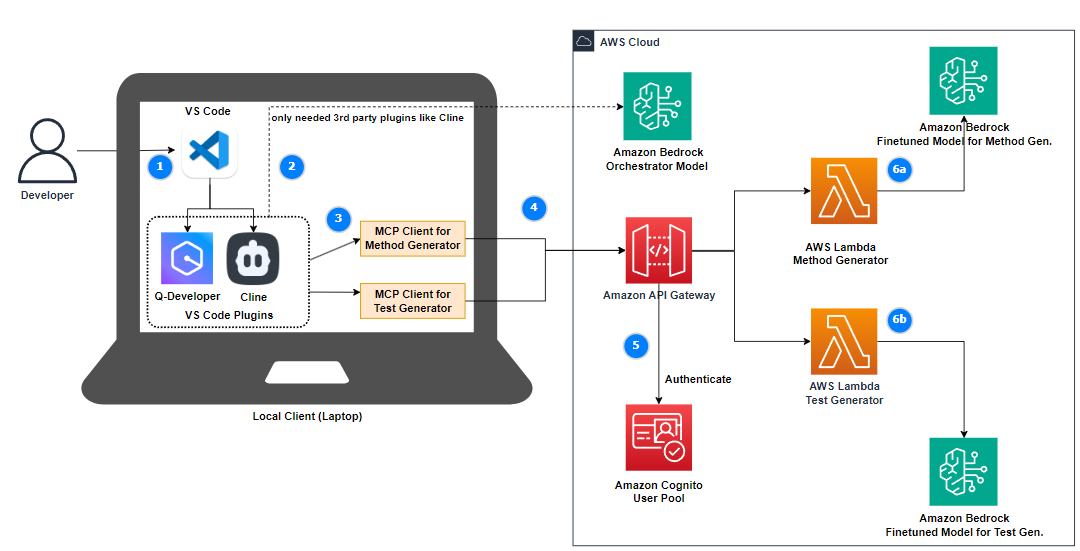}
\caption{System architecture. The MCP server exposes three tools (\code{generate}, \code{edit}, \code{verify}) that AI coding assistants invoke. Requests route through API Gateway to models on Amazon Bedrock or SageMaker.} \label{fig_architecture}
\end{figure*}

Our architecture (Figure \ref{fig_architecture}) integrates with existing development environments while maintaining separation between orchestration and specialized generation models. This separation is what makes the assistant agentic: the developer works in an IDE plugin such as Cline, whose orchestrator model decides which MCP tool a task requires and dispatches the request rather than answering from its own weights. Requests route through API Gateway, authenticated via Cognito, to Lambda functions that invoke the models we host for the automotive domain, so the orchestrator sequences the calls while specialized models generate code.

\section{Evaluation Setup}\label{sec:evaluation}
Evaluating automotive code quality requires both automated metrics and human expertise. Table \ref{tab:evaluation_metrics} in Appendix summarizes candidate metrics by type, tool availability, complexity, reliability, and cost, spanning reference-based similarity measures \citep{paul2024benchmarks,takaichi2022nlp}, static and dynamic analysis, and standards conformance against MISRA \citep{motor2013misra}, CERT C \citep{seacord2008cert}, and AUTOSAR \citep{autosar2017guidelines}. We do not adopt the similarity measures as primary metrics, since closeness to a reference implementation is a weak proxy for the standards adherence developers must certify.

Our evaluation covers three dimensions. First, we score code against the rule sets of Section~\ref{subsec:rules}, comparing generated functions to a human baseline of production code written before the assistant was introduced on 822 held-out functions (574 C, 248 C++). Second, we measure generation latency. Third, we pilot the assistant with automotive engineers and survey their usage patterns.

To score code automatically, we implement rule sets as pattern-based (regular-expression) static checkers. The rules they encode are drawn from verification patterns AUMOVIO already applies\footnote{Licenses are required for the use of these standards.}, selected by engineers as high-impact patterns. They are a deliberately simplified subset of AUMOVIO's verification process: rules requiring semantic, data-flow, or inter-procedural reasoning are excluded by construction, since a pattern matcher cannot decide them and may produce false positives and false negatives. Production use therefore requires established automotive linters.

\section{Results}\label{sec:results}
\subsection{Pattern Conformance Evaluation}

Table \ref{tab:model_comparison} reports mean pattern-level rule violations per generated function alongside the observed end-to-end generation latency in seconds.

Holding the base model fixed at Qwen3 32B, PE gives the fewest C violations on both checkers (1.22 coding, 0.54 architecture), ahead of either FT variant. On C++ the ordering reverses, with FT on curated data recording the fewest coding (1.25) and architecture (0.02) violations. These margins are thin: the architecture column separates all three Qwen variants by at most 0.08 violations per function, which we do not treat as a meaningful difference. The same curated variant is also the worst Qwen variant on both C checkers (1.67 each), so its C++ gain comes with a C cost rather than a uniform improvement. Across model families, Claude Sonnet 4 (PE) and Amazon Nova Pro (PE) produce more violations in both languages than the best Qwen3 32B variant.

One plausible reading of the C versus C++ divergence is that the two rule families differ in locality. Our C rules mostly describe function-local obligations, such as validating a pointer before dereferencing it or checking buffer bounds, which can be stated compactly in a prompt and applied by a general-purpose model without prior exposure to the codebase. The C++ architecture rules instead encode the AUMOVIO's layering and interface conventions, which are longer-range, idiosyncratic to their projects, and largely absent from pretraining data. These are difficult to convey exhaustively in a prompt but appear repeatedly in training examples. If this reading is right, fine-tuning helps less by teaching C++ syntax than by absorbing organization-specific convention.

Observed generation latency ranges from ${\sim}$15 seconds (Nova Pro, PE) to ${\sim}$110 seconds (Qwen3 32B, PE). Note that these figures are not a controlled comparison of the customization strategies, since PE models are served through a managed API while FT models run on separately provisioned inference infrastructure. At the same time, the numbers characterize what developers in our pilot actually waited for when using the assistant. The spread is wide enough to motivate hybrid serving, with faster models reserved for interactive editing and slower ones for long-form function generation.

\begin{table*}[t]\centering
\caption{Automotive code generation: mean \emph{pattern-level} rule violations per function (lower is better) and observed end-to-end generation latency. Counts reflect conformance to patterns our custom checkers can match, not MISRA/AUTOSAR compliance or functional correctness. Models are evaluated on 822 test examples (574 C, 248 C++) from production automotive code. \textbf{Bold} = best per column; \underline{underline} = worst per column.}
\label{tab:model_comparison}
\small
\begin{tabular*}{\textwidth}{@{\extracolsep{\fill}}lccccr@{}}
\toprule
\textbf{Model} & \textbf{\begin{tabular}[c]{@{}c@{}}Coding\\patterns (C)\end{tabular}} & \textbf{\begin{tabular}[c]{@{}c@{}}Architecture\\patterns (C)\end{tabular}} & \textbf{\begin{tabular}[c]{@{}c@{}}Coding\\patterns (C++)\end{tabular}} & \textbf{\begin{tabular}[c]{@{}c@{}}Architecture\\patterns (C++)\end{tabular}} & \textbf{\begin{tabular}[c]{@{}c@{}}Generation\\latency (s)\end{tabular}} \\
\midrule
Qwen3 32B (PE) & \textbf{1.22} & \textbf{0.54} & 1.73 & 0.10 & \underline{110.19} \\
\addlinespace
Qwen3 32B (FT) & 1.31 & 0.84 & 1.81 & 0.09 & 93.69 \\
\addlinespace
Qwen3 32B (FT, curated) & 1.67 & 1.67 & 1.25 & \textbf{0.02} & 85.34 \\
\addlinespace
\midrule
\addlinespace
Sonnet 4 (PE) & 1.80 & 1.84 & 2.61 & 0.12 & 28.68 \\
\addlinespace
Nova Pro (PE) & 1.63 & 1.48 & \underline{2.88} & \underline{0.24} & \textbf{14.62} \\
\addlinespace
\midrule
\addlinespace
Human code (production) & \underline{2.45} & \underline{2.20} & \textbf{0.84} & 0.06 & N/A \\
\bottomrule
\end{tabular*}
\begin{flushleft}\footnotesize
\end{flushleft}
\end{table*}

\subsection{Comparison with the Human Baseline}\label{subsec:gameable}

Human production code records 2.45 and 2.20 violations per function on the two C checkers, worse than every generated variant. Read naively, LLM-generated C code adheres to automotive rules better than code written by the AUMOVIO's own engineers and shipped in production vehicles. However, inspecting the outputs hints why this reading does not hold: generated functions apply defensive constructs, with null checks on pointer parameters, explicit bounds checks, single-exit control flow, and explicit error returns, and our checkers reward these surface patterns. Human code frequently omits a local null check because the invariant holds elsewhere, at a caller or through a module-level precondition that the author knows to hold and that a pattern matcher cannot see. The checker counts an absent local check as a violation whether or not the property holds non-locally, so it penalizes code that reasons beyond the function boundary and rewards code that re-checks everything. Such redundancy is not free in embedded settings, where extra branches cost code space and complicate timing analysis.

The C++ results point the other way, with human code best on coding patterns (0.84). Our C++ rule sets weight architectural and interface conventions more heavily than local defensive checks, so uniform guard insertion earns less credit there, while the AUMOVIO's own code follows conventions its authors apply routinely. Together, the results suggest the checkers measure how closely code resembles a pattern vocabulary, which generated code matches well when the rules are local and less well when they encode organization-specific structure. The problem is that a pattern-matching proxy for a standard can be optimized without being satisfied: a model raises its score by emitting recognized constructs rather than by producing safer code. Such checkers are therefore limited but are still useful as a fast pre-commit filter and as generation context.

\subsection{Pilot Study with Developers}\label{subsec:pilot}

Pattern conformance metrics describe what the models can produce in isolated tasks. To test the actual human-agent collaboration, we run a two-week pilot with nine of the AUMOVIO's developers and analyze post-usage survey results.

Pilot participants are automotive engineers on AUMOVIO's active C/C++ projects, recruited voluntarily from teams scheduled to trial AI assistance. Each installs the MCP server into Cline and works on their own production tasks rather than assigned benchmarks. All three tools are exposed through MCP descriptions, so the assistant can select them autonomously, and \code{verify} is additionally documented as directly invocable.

Evidence comes from a concluding survey covering experience, usage frequency, tool usage, usefulness, and other aspects. Every quantity is self-reported, since we collect no telemetry on invocation counts. Note that with nine respondents, we read only large gaps in the results as informative.

\begin{figure*}[t]\centering
\includegraphics[width=0.88\textwidth]{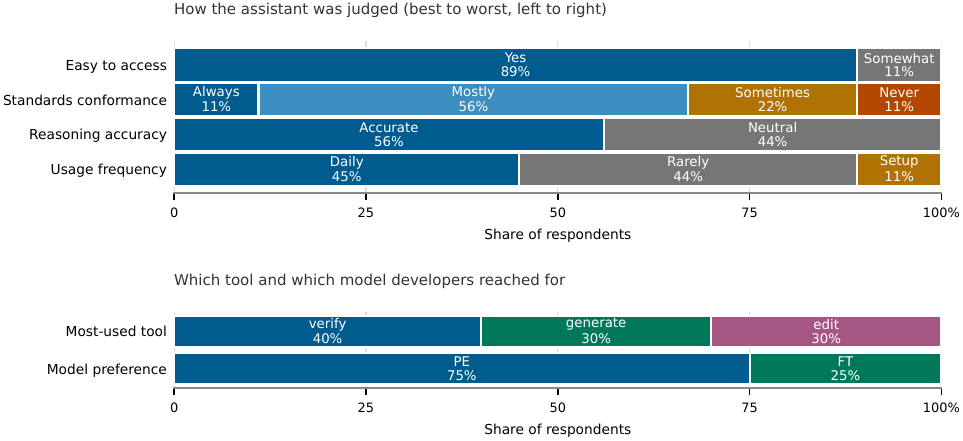}
\caption{Pilot survey response distributions ($n{=}9$ developers). The upper panel holds judgments of the assistant, ordered from most to least favorable; the lower panel holds choices among tools and modes. Denominators differ by item, since not every respondent answered every question; raw counts are given in the text. Only large gaps are informative and percentages are rounded to sum to 100. ``Setup'' denotes participants who installed the assistant but reported no working use. All responses are self-reported.}
\label{fig:pilot_survey}
\end{figure*}

Developers rate their overall experience with the assistant 3.89 out of 5, and their willingness to recommend it to colleagues 4.00 out of 5. On average, the productivity improvements is marked at 4.67 on a seven-point scale. The distributions behind these means are more informative: no respondent gives the lowest rating on any item, but none is unanimous: experience draws 3 top ratings vs 1 low, recommendation -- 4 top against 3 at or below the midpoint Figure \ref{fig:pilot_survey} provides more detail on usability: 8 of 9 respondents find the assistant easy to integrate, and 5 of 9 rate its reasoning accurate against 4 neutral and none inaccurate. Usage does not settle at a middle value: four of nine respondents use the assistant daily and four use it rarely, forming two distinct user populations.

Three patterns stand out in the survey results. First, 6 of the 8 respondents who answered this question prefer the PE mode. The PE configurations served in the pilot include the API-hosted models, which answer in roughly 15 to 29 seconds against 85 to 94 for the FT models (Table~\ref{tab:model_comparison}), so mode preference and response time are confounded here. Combined with developer feedback, we read this as evidence that in tight read-edit-review loops generation latency behaves as a key property.

Second, 6 of 9 developers report conformance \emph{always} or \emph{most of the time}, but only one of them answers \emph{always}, while 2 answer \emph{sometimes} and 1 \emph{never}. This matters because in Table~\ref{tab:model_comparison} the same models score at or better than human code on rule set conformance. Developers do not internalize the favorable automated numbers as trust.

Third, \code{verify} draws the largest share of most-used-tool responses (4 of 10 selections, against 3 each for \code{generate} and \code{edit}; the item permitted multiple selections) and the largest \emph{very useful} share. Respondents also run \code{verify} on existing human-written code, not only on generated output, a use a generation-only assistant would not support.

\section{Conclusion}\label{sec:conclusion_further_work}
AI assistance helps safety-critical development only if built around how such code is accepted: reviewed, traced to a requirement, checked before commit. We present an MCP-based coding assistant addressing this by: (i) treating verification as a peer of generation rather than a step hidden inside it; (ii) adapting the model to the domain through prompt engineering and fine-tuning on production code paired with the requirements each function must satisfy. We deploy it on AUMOVIO's active automotive C/C++ projects, evaluate pattern conformance, and run a developer pilot in production.

Our experiments produce three findings. First, \code{verify} is the most-used tool in the pilot, and developers run it on their own existing code rather than only on generated output, which indicates that verification is used as a first-class step. Second, with the base model held fixed, PE gives fewer C violations while the C++ counts separate the two strategies by margins too small to call, so we do not read either as the better choice. Third, developers prefer the faster configuration and do not fully trust the output despite its scores.

Together, these results suggest that coding assistants for safety-critical domains should be judged on more than generation quality. What mattered to our developers was checking code themselves, getting answers fast enough to stay in a review loop, and knowing where the automated verdict held.

\section*{Acknowledgments}
We thank the wider AUMOVIO team who supported us throughout the project. We are especially grateful to Stefan-Cosmin Grigoriu for his support.

\section*{Limitations}

Our evaluation has several limitations. Section~\ref{sec:evaluation} states what our pattern checkers can and cannot show, and Section~\ref{subsec:gameable} shows they favor generated code. We also report no functional correctness: no compilation or unit-test pass rates, and no check that a function satisfies the requirements or UML behavior in its prompt.

On the model comparison, we use parameter-efficient LoRA fine-tuning rather than full fine-tuning, and run no ablations over prompt design, context elements, or rule-set inclusion. We report no significance tests, so the small C++ margins should not be read as ranking the strategies. Latency was measured on different serving stacks, so it describes our deployment rather than the general strategies.

On the human evidence, the pilot has nine self-selected developers working with the assistant for two weeks. Everything is self-reported, with no telemetry and no record of whether generated code was committed or how much it was edited first, so the results are directional signals rather than effect estimates. Finally, all data comes from one organization, and the code, rule sets, and checkers are proprietary and cannot be released publicly.

\section*{Ethical Considerations}

Defective code in the automotive setting can harm people, so the assistant is not meant to automate compliance work: everything it produces needs review by qualified engineers. Three failure modes matter in deployment. \textit{False confidence}, where code that passes pattern checks still violates functional or timing requirements; Section~\ref{subsec:gameable} is a concrete case, since adding defensive patterns everywhere raises the score without making the code safer, and the extra branches can themselves affect code size and worst-case execution time. \textit{Correlated blind spots}, where the generator and the checker share a model family and miss the same defect twice, which is why we keep the checker independent of the generator and why extending the architecture to test generation would need the same separation. \textit{Conformance drift}, where non-conforming code spreads through reuse. Human approval gates, keeping the checker independent of the generator, and logging prompts, context, model IDs, and analyzer reports all help.

We ran no red-teaming and did not test whether the assistant can be pushed into producing insecure code, tricked into evading the checkers, or made to introduce vulnerabilities our rules do not cover. Production readiness would require that work.

The study uses AUMOVIO's proprietary code under their data governance. Pilot participation was voluntary, and we collected no personal data beyond anonymized survey responses.

\bibliography{references}

\clearpage
\appendix

\section{Evaluation Metrics Overview}\label{app:metrics}

Choosing what to measure is itself a decision when evaluating safety-critical automotive code, since no single metric captures its quality and the practical options differ in how much tooling they need and how much their verdicts can be trusted. Table~\ref{tab:evaluation_metrics} surveys the metric families we considered, rating each on tool support, implementation complexity, reliability, and cost. The comparison motivates the choices in Section~\ref{sec:evaluation}: we rely on standards conformance, which combines high tool support with high reliability, and on support metrics such as latency, while declining reference-based similarity measures whose reliability is only moderate and whose relevance to certification is weak.

\begin{table*}[b]\centering
\caption{Evaluation metrics for automotive code quality assessment. Each category is rated on Tool Support (availability of automated tools), Complexity (implementation effort), Reliability (consistency of results), and Cost (resources required). High = better, except for Cost and Complexity where low is preferable.}
\label{tab:evaluation_metrics}
\footnotesize
\begin{tabular*}{\textwidth}{@{\extracolsep{\fill}}lccccc@{}}
\toprule
\textbf{Category} & \textbf{Example Metrics} & \textbf{Tool Support} & \textbf{Complexity} & \textbf{Reliability} & \textbf{Cost} \\
\midrule
ML Static Analysis & BLEU, ROUGE, METEOR & High & Medium & Medium & Low \\
\addlinespace
Static Analysis & Linters, Complexity & High & Low & High & Medium \\
\addlinespace
Dynamic Analysis & Coverage, Runtime & Medium & High & High & High \\
\addlinespace
Standards Compliance & Industry Coding Standards & High & Medium & High & Medium \\
\addlinespace
 & (e.g., MISRA, CERT C) &  &  &  &  \\
\addlinespace
Support Metrics & Latency, Time & Low & Low & Medium & Low \\
\addlinespace
Business KPIs & ROI, Time-to-Market, Productivity & Low & Low & Low & Medium \\
\bottomrule
\end{tabular*}
\end{table*}

\section{Dataset Preprocessing Pipeline}\label{app:preprocessing}

This appendix expands the pipeline summarized in Section~\ref{subsec:dp}. The preprocessing pipeline begins by traversing automotive C/C++ repositories to extract individual functions along with their surrounding context, which includes:

\begin{itemize}
    \item \textbf{Function Documentation}: Both Doxygen-style comments \citep{doxygen} and inline documentation are extracted and linked to their corresponding function implementations
    \item \textbf{System Requirements}: The pipeline parses DOORS export XML files \citep{ibm_doors} to map requirement identifiers mentioned in function documentation to their full requirement texts
    \item \textbf{Architectural Context}: PlantUML diagrams \citep{plantuml} referenced in documentation are extracted and included to provide behavioral specifications
    \item \textbf{Code API Context}: Related header files and their function signatures are gathered to provide information about available APIs and data structures
\end{itemize}

Header linking works as follows: the system identifies the main header file corresponding to each C/C++ source file and extracts additional context from included dependencies, so that generated code can reference existing APIs rather than reimplementing them.

The context aggregation process is implemented as follows:

\begin{lstlisting}[language=Python, caption=Context aggregation from preprocessing pipeline,  basicstyle=\ttfamily\scriptsize, label=processing_listing, breaklines=true, breakatwhitespace=false, postbreak=\mbox{\textcolor{red}{$\hookrightarrow$}\space}, xleftmargin=2.5em, framexleftmargin=2.5em]
def create_training_example(function_info):
    user_message = f"Implement the function: {function_info['signature']}\n\n"

    if function_info['documentation']:
        user_message += f"with following specifications:\n{function_info['documentation']}"

    if function_info['requirements']:
        user_message += f"\n\nRequirements texts:\n{function_info['requirements']}"

    if function_info['uml_diagram']:
        user_message += f"\n\nThe behavior follows this UML diagram:\n{function_info['uml_diagram']}"

    return {
        "messages": [
            {"role": "user", "content": user_message},
            {"role": "assistant", "content": function_info['implementation']}
        ]
    }
\end{lstlisting}

For quality filtering and validation, the preprocessing pipeline implements several quality assurance mechanisms:

\begin{itemize}
    \item \textbf{Function Signature Validation}: Automatically corrects function signatures in implementation files by matching them against header file declarations
    \item \textbf{Documentation Completeness}: Only functions whose documentation specifies the intended behavior are included in the training set
    \item \textbf{Code Compliance}: Functions are validated against AUTOSAR \citep{autosar2017guidelines} and MISRA-C \citep{motor2013misra} patterns to assess the initial compliance level
\end{itemize}

Finally, to ensure balanced representation across different code complexity levels, the preprocessing pipeline implements stratified sampling based on function length and complexity. This approach creates training and test sets that maintain consistent distribution characteristics. 
The resulting dataset contains approximately 6,000 high-quality function implementations with their complete contextual information, split into training and evaluation sets while maintaining complexity distribution balance.

\section{Model Architecture and Training Configuration}\label{app:training}

We select Qwen3-32B as the base model \citep{yang2025qwen3technicalreport} for its performance on code generation tasks at a size we could fine-tune on available hardware. Training uses Low-Rank Adaptation \citep{hu2021loralowrankadaptationlarge} with rank-8 adapters and $\alpha=16$, applied to the query, key, value, and output projections together with all feed-forward components, plus 4-bit quantization via BitsAndBytes \citep{dettmers2023qlora} to reduce the memory footprint. None of these choices are specific to automotive code: they are a standard parameter-efficient recipe, reported here for reproducibility rather than offered as contributions.

Training runs on Amazon SageMaker with PyTorch DeepSpeed \citep{rasley2020deepspeed}, orchestrated across multiple GPUs through SageMaker's remote function decorator, as shown in Listing~\ref{sagemaker_listing}.

\begin{lstlisting}[language=Python, caption=SageMaker remote training configuration, label=sagemaker_listing, basicstyle=\ttfamily\scriptsize, breaklines=true, breakatwhitespace=false, postbreak=\mbox{\textcolor{red}{$\hookrightarrow$}\space}, xleftmargin=2.5em, framexleftmargin=2.5em]
@remote(
    instance_type="ml.p4d.24xlarge",
    volume_size=100,
    use_torchrun=True,
    pre_execution_commands=[
        "pip install torch==2.5.1 transformers==4.51.3",
        "pip install peft==0.15.2 deepspeed bitsandbytes",
    ]
)
def train_model(train_dataset, test_dataset, config):
    # Training implementation with adaptive DeepSpeed configuration
    stage = 2 if use_quantization else 3
    deepspeed_config = {
        "zero_optimization": {
            "stage": stage,
            "overlap_comm": True,
            "contiguous_gradients": True,
            "offload_optimizer": {"device": "cpu", "pin_memory": True}
        }
    }
\end{lstlisting}

We select the DeepSpeed ZeRO stage \citep{rajbhandari2020zeromemoryoptimizationstraining} according to whether quantization is enabled. ZeRO-2 partitions optimizer states and gradients while replicating parameters on each device; ZeRO-3 additionally partitions the parameters, which permits larger models at the cost of communication overhead and, for our long prompts, distributes the memory burden of storing full parameters. We default to ZeRO-2 under quantization, since quantized parameters interact poorly with ZeRO-3's parameter-gathering, and ZeRO-2 also requires fewer all-gather operations. The remaining settings are conventional: gradient accumulation for a larger effective batch size, gradient checkpointing and CPU offloading to fit the model in memory, Flash Attention 2 \citep{dao2022flashattentionfastmemoryefficientexact, dao2023flashattention2fasterattentionbetter}, and NCCL with communication overlap and extended timeouts. The one detail worth flagging is that our prompts are long, since they carry resolved requirement texts, UML fragments, and header context, so context length rather than model size drove most of our memory pressure.

The final model is packaged through Amazon Bedrock's custom model import functionality for integration with the multi-model architecture. The distributed setup reaches approximately 85\% GPU utilization with stable convergence, which kept each fine-tuning cycle within a development sprint.

\section{Conversation Format Example}\label{app:format}

The training data utilizes HuggingFace's chat template \citep{hf_chat_tmplte} to structure the automotive coding task as a conversational interaction. Listing~\ref{conv_listing} gives the full, untruncated version of the example abbreviated as Listing~\ref{code_example} in Section~\ref{subsec:dp}, showing both resolved requirement texts, the complete PlantUML fragment, and the target implementation as it appears in training. This conversational format supplies the model with the requirements, safety level, and behavioral specification alongside the signature, so that the obligations a function must satisfy are present in the prompt rather than inferred from the signature alone.

\begin{lstlisting}[language=Python, caption=HuggingFace chat template example for automotive C++ code, basicstyle=\ttfamily\scriptsize, label=conv_listing, breaklines=true, breakatwhitespace=false, postbreak=\mbox{\textcolor{red}{$\hookrightarrow$}\space}, xleftmargin=2.5em, framexleftmargin=2.5em]
training_example = {
    "messages": [
        {
            "role": "user", 
            "content": """Implement the function: 
void ValidateEngineParameters(const EngineData_t* engineData, ValidationResult_t* result)

with following specifications:
/**
 * @brief Validates engine parameters according to AUTOSAR specification
 * @param engineData Pointer to engine data structure containing RPM, temperature, pressure
 * @param result Pointer to validation result structure
 * @requirements REQ_ENG_001: Engine RPM must be within 800-6000 range
 * @requirements REQ_ENG_002: Engine temperature must not exceed 110 degrees C
 * @safety_level ASIL-D
 */

Requirements texts:
REQ_ENG_001: The engine control unit shall validate that engine RPM remains within the 
operational range of 800 to 6000 RPM during normal operation.
REQ_ENG_002: The engine control unit shall monitor engine coolant temperature and trigger
protective measures when temperature exceeds 110 degrees C.

The behavior follows this UML diagram:
@startuml
ValidateEngineParameters -> CheckRPMRange : validate RPM
CheckRPMRange -> CheckTemperature : if RPM valid
CheckTemperature -> SetValidationResult : set result
@enduml"""
        },
        {
            "role": "assistant",
            "content": """
void ValidateEngineParameters(const EngineData_t* engineData, ValidationResult_t* result)
{
    /* MISRA-C 2012 Rule 1.3: No undefined behavior */
    if ((engineData == NULL) || (result == NULL))
    {
        if (result != NULL)
        {
            result->isValid = false;
            result->errorCode = VALIDATION_NULL_POINTER;
        }
        return;
    }

    /* Initialize result structure */
    result->isValid = true;
    result->errorCode = VALIDATION_SUCCESS;

    /* REQ_ENG_001: Validate RPM range (800-6000) */
    if ((engineData->rpm < 800U) || (engineData->rpm > 6000U))
    {
        result->isValid = false;
        result->errorCode = VALIDATION_RPM_OUT_OF_RANGE;
        return;
    }

    /* REQ_ENG_002: Validate temperature limit (110 degrees C) */
    if (engineData->temperature > 110.0f)
    {
        result->isValid = false;
        result->errorCode = VALIDATION_TEMPERATURE_EXCEEDED;
        return;
    }
    /* All validations passed (result pre-initialized to success) */
}"""
        }
    ]
}
\end{lstlisting}

\end{document}